\documentclass[a4paper,10pt]{article}

\newcommand{\vs}[1]{\rule[- #1 mm]{0mm}{#1 mm}}

\newcommand{\lbl}[1]{\label{eq:#1}}
\newcommand{\rf}[1]{(\ref{eq:#1})}
\newcommand{\nn}{\nonumber}

\newcommand{\be}{\vs{2}\begin{equation}}
\newcommand{\ee}{\vs{2}\end{equation}}

\newcommand{\bea}{\begin{eqnarray}}
\newcommand{\ena}{\end{eqnarray}}
\newcommand{\nnbea}{\begin{eqnarray*}}
\newcommand{\nnena}{\end{eqnarray*}}

\newcommand{\lra}{\ \longrightarrow\ }

\newcommand{\xx}{(x)}
\newcommand{\yy}{(y)}

\newcommand{\cR}{{\cal R }}
\newcommand{\cW}{{\cal W }}

\newcommand{\cK}{{\cal K }}

\newcommand{\cC}{{\cal C }}

\newcommand{\cG}{{\cal G }}
\newcommand{\cD}{{\cal D }}
\newcommand{\cL}{{\cal L }}

\newcommand{\cM}{{\cal M }}
\newcommand{\cS}{{\cal S }}

\newcommand{\cX}{{\cal X }}

\newcommand{\prt}{\partial}

\newcommand{\zer}[1]{\stackrel{\circ}{#1}}

\newtheorem{Results}{Results}[section]
\newcount\a
\newcount\b
\def\dayofweek{
  \a=\year\advance\a by -1980\b=\a
  \advance\b by 3
  \ifnum\month>2\advance\b by 1\fi
  \divide\b by 4             % Number of Leap Days passed
  \advance\a by \b           % Number of days passed.
  \advance\a by \ifcase\month
     \or1\or32\or60\or91\or121\or152%
     \or182\or213\or244\or274\or305\or335\fi
  \advance\a by \day
  \b=\a \divide\b by 7 \multiply\b by 7 \advance\a by-\b
  \ifcase\a Sunday\or Monday\or Tuesday\or
    Wednesday\or Thursday\or Friday \or Saturday\fi
  }

\def\today{\ifcase\month\or
  January\or February\or March\or April\or May\or June\or
  July\or August\or September\or October\or November\or December\fi
  \space\number\day, \number\year}

\begin{document}

%%%%%%%%%%%%%%%%%%%%%%
%\date{\today}
%\renewcommand{\thefootnote}{\fnsymbol{footnote}}
%\setcounter{page}{1}

%----------------------------------------------------------------

%\begin{frontmatter}
%\year=2002                                                                                       \begin{scriptsize} \begin{normalsize} \begin{large}  \end{large} \end{normalsize} \end{scriptsize}
\vskip -1.5cm
%\begin{eqnarray}
\title{\bf {On an  extension of General Coordinate Transformations Algebra. }}
\vskip -1.5cm

\author{ Giuseppe Bandelloni% $^a$%\\[6mm]
 \\[6mm]
%$^a$
Dipartimento di Fisica
dell'Universit\`a di Genova,\\
 Via Dodecaneso 33, I-16146 GENOVA-Italy\\
 and \\
Istituto Nazionale di Fisica Nucleare, INFN, Sezione di Genova\\
via Dodecaneso 33, I-16146 GENOVA Italy\\
e-mail : {\tt beppe@genova.infn.it}\\[4mm]
}

%\\[6mm]
\maketitle
%\vskip -1.5cm
%\begin{titlepage}
%\title{LargeDiff}
%\author{    ......  }
%\date{       \today-\dayofweek}

%\end{titlepage}
%\maketitle
\abstract{ An   extension of the General Coordinate
Transformations algebra is constructed by means geometrical
consistency conditions. An class of infinite invariants is
derived. In particular we construct the consistent extension of
the gravitational anomaly for each even dimension. The new
contributions for these anomalies allow to define an improved
Ward operator for which the symmetry is restored.}

\indent

\section{Introduction}
\lbl{introduction}

 Over the course of the last century it has become clear
that both elementary particle physics and relativity theories
are based on the notion of symmetries. These symmetries become
manifest in that the "laws of nature" are invariant under
spacetime transformations and/or gauge transformations. In fact
the dynamics of physical theories is deeply intertwined with
space-time reparametrization invariance; indeed it assures the
independence of the physics from the frames, and so endorses
the universality requirements. These fulfillments can be
assured in several ways, the most canonical ones are performed
or via finite reparametrizations (change of charts) or  through
infinitesimal transformations,
 which allow the use of the General Coordinate Transformations  (diffeomorphism) algebra within the most common treatments  of the physical theories (see Ref\cite{DeWitt:1965jb} for an introduction)

Moreover one can find in the Literature a widespread of intermediate approaches; we recall the use of discrete coordinate transformation \cite{Regge:1961px},\cite{Loll:1998aj} , but we focus the attention to the algebra extension methods, which, in the past Literature found a great popularity within the holomorphic two dimensional models with the discovery of the so-called $\cW$-algebras \cite{Zamolodchikov:1985wn} \cite{DiFrancesco:1990qr}.
This approach appears, after many years of investigations, as a clever improvement of the infinitesimal transformations approach.
 So their extension to higher dimensions might be interesting in those cases where the smoothness of the
 coordinates change is not verified.
    We shall perform the extensions via the B.R.S. \cite{Becchi:1975nq} consistency requirements,
    inserting first the infinitesimal contributions, and then adding new further terms,
    which define the extension, whose algebraic properties will be defined by the nilpotency of
     the algebra transformations.
     It is easy to realize that this algebra  process is not a simple task,
     and a priori depends on the space-time geometries, so that  the roads covered by
      a well defined deformation are not generally straight, but their  tortuosity
      have to be settled by gates defined be precise algebraic supplementary conditions.
      So we can expect that this treatment will have repercussions on the physical landscape.
       This can be manifested by the Lagrangians and Anomalies changes. It would be interesting if  this
       new expansion  would produce new
welcome   physics perspectives.

\vskip0.5 cm

In Section \rf{algebraextension} we shall perform an infinite
expansion, whose consistency constraints greatly reduce the
B.R.S. ghosts from infinity to two, with well defined
geometrical properties.

\vskip 0.5 cm

Then in Section \rf{anomalies} we calculate the anomalies
change. This improvement reveals a welcome intertwining
property among the ghosts, which, at any dimension, permits to
define a combined, anomaly free, Ward operator.

\vskip 0.5 cm

Few conclusions are reported in Section \rf{conclusions}, and
in the Appendix \rf{appendix} we report the  calculations which
lead to the explicit expressions of the anomalies at any
dimensions quoted in Section \rf{anomalies}

\vskip 1.0 cm

\section{  The diffeomorphism algebra extension in a B.R.S. approach}
\lbl{algebraextension}
 Our treatment begins  supposing that the Riemmanian connection $\Gamma_{(\mu,\lambda)}^\nu\xx$, due to an improvement, satisfies the infinitesimal (B.R.S ) extended reparametrization transformation:
\begin{eqnarray}
&&\cS\Gamma_{(\mu,\lambda)}^\nu\xx=\cC^\sigma\xx\prt_\sigma\Gamma_{(\mu,\lambda)}^\nu\xx+\prt_\mu\cC^\sigma
\Gamma_{(\sigma,\lambda)}^\nu\xx+\prt_\lambda\cC^\sigma\xx\Gamma_{(\mu,\sigma)}^\nu\xx
-\prt_\sigma\cC^\nu\xx\Gamma_{(\mu,\lambda)}^\sigma\xx-\prt_\mu\prt_\lambda\cC^\sigma\xx+
\Omega_{(\mu,\lambda)}^\nu\xx\nn\\
\lbl{sgamma}
\end{eqnarray}
where $\Omega_{(\mu,\lambda)}^\nu\xx$ is a new (general) tensorial extension term
(absent in the ordinary general coordinate transformations algebra), which could reproduce
 higher diffeomorphisms corrections, and $\cC^\mu\xx$ is the usual diffeomorphism B.R.S. ghost.

Equation \rf{sgamma} has a full covariant rewriting, in terms of the Riemmaniann curvature $\cR^{(\nu)}_{([\rho,\lambda],\mu)}\xx$, as:
\begin{eqnarray}
\cS \Gamma_{(\mu,\lambda)}^\nu\xx &=& -\cD_\lambda\cD_\mu\cC^\nu\xx+
\cC^{(\rho)}\xx\cR^{(\nu)}_{([\rho,\lambda],\mu)}\xx+\Omega_{(\lambda,\mu)}^\nu\xx\nn\\
\lbl{sgammacov}
\end{eqnarray}
where $\cD_\lambda$ indicates the the covariant derivative.

In this spirit the ghost $\cC^\mu\xx$ must undergo a new extended transformation:
\begin{eqnarray}
\cS \cC^\mu\xx &=&  \cC^\lambda\xx \cD_\lambda \cC^\mu\xx +\cX^\mu\xx\nn\\
% \cS \Gamma_{(\mu,\nu)}^\lambda\xx &=& -\cD_\mu\cD_\nu\cC^\lambda\xx+\Sigma_{(\mu,\nu)}^\lambda\xx
\lbl{sghost}
 \end{eqnarray}
So,we have a new law for the B.R.S curvature transformation:
\begin{eqnarray}
\cS\cR_{([\nu,\mu],\rho)}^\sigma\xx&=&\cD_\nu\biggl(\cS \Gamma_{(\mu,\rho)}^\sigma\xx\biggr)-\cD_\mu\biggl(\cS \Gamma_{(\nu,\rho)}^\sigma\xx\biggr)
\nn\\
&=&\cR_{([\mu,\nu]\rho)}^\lambda\xx\cD_\lambda\cC^\sigma\xx-\cR_{([\mu,\nu],\lambda)}^\sigma\xx
\cD_\rho\cC^\lambda\xx+
\cD_\nu\cC^\lambda\xx\cR_{([\lambda,\mu],\rho)}^\sigma\xx\nn\\
&+&
\cC^\lambda\xx\cD_\nu\cR_{([\lambda,\mu],\rho)}^\sigma\xx-
\cD_\mu\cC^\lambda\cR_{([\lambda,\nu]\rho)}^\sigma\xx-\cC^\lambda\xx
\cD_\mu\cR_{([\lambda,\nu]\rho)}^\sigma\xx\nn\\
&+&
\cD_\nu\Omega_{(\mu,\rho)}^\sigma\xx-\cD_\mu\Omega_{(\nu,\rho)}^\sigma\xx\nn\\
\lbl{sR}
\end{eqnarray}
In the B.R.S. formalism the  consistency is fulfilled by the
algebra nilpotency requirements, which give for the connection extensions the B.R.S. full covariant transformations:
%\begin{eqnarray}
%&&\cS\Omega_{(\mu,\lambda)}^\nu\xx=\cC^\sigma\xx\prt_\sigma\Omega_{(\mu,\lambda)}^\nu\xx+\prt_\mu\cC^\sigma
%\Omega_{(\sigma,\lambda)}^\nu\xx+\prt_\lambda\cC^\sigma\xx\Omega_{(\mu,\sigma)}^\nu\xx
%-\prt_\sigma\cC^\nu\xx\Omega_{(\mu,\lambda)}^\sigma\xx\nn\\&-&
%\biggl(\cX^\sigma\xx\prt_\sigma\Gamma_{(\mu,\lambda)}^\nu\xx+\prt_\mu\cX^\sigma
%\Gamma_{(\sigma,\lambda)}^\nu\xx+\prt_\lambda\cX^\sigma\xx\Gamma_{(\mu,\sigma)}^\nu\xx
%-\prt_\sigma\cX^\nu\xx\Gamma_{(\mu,\lambda)}^\sigma\xx-\prt_\mu\prt_\lambda\cX^\sigma\xx\biggr)\nn\\
%\lbl{somega}
%\end{eqnarray}

%The previous equation \rf{somega} can be rewritten in a full covariant way as:
\begin{eqnarray}
&&\cS\Omega_{(\mu,\lambda)}^\nu\xx=\cC^\sigma\xx\cD_\sigma\Omega_{(\mu,\lambda)}^\nu\xx
+\cD_\mu\cC^\sigma
\Omega_{(\sigma,\lambda)}^\nu\xx\nn\\
&&+\cD_\lambda\cC^\sigma\xx\Omega_{(\mu,\sigma)}^\nu\xx
-\cD_\sigma\cC^\nu\xx\Omega_{(\mu,\lambda)}^\sigma\xx+\cD_\mu\cD_\lambda\cX^\nu\xx-\cX^\sigma\xx
\cR_{([\sigma,\mu],\lambda)}^\nu\xx\nn\\
\lbl{somegacov}
\end{eqnarray}

On the other hand. the ghost algebra extension $\cX^\sigma \xx$
has (due to the nilpotency of Equation \rf{sghost}) a B.R.S.
transformation:
\begin{eqnarray}
\cS\cX^\sigma\xx
&=&\cC^\lambda\xx\prt_\lambda\cX^\sigma\xx-\prt_\lambda\cC^\sigma\xx\cX^\lambda\xx\nn\\
\lbl{SX}
\end{eqnarray}

It is trivial to verify $\cS^2\Omega_{(\mu,\lambda)}^\nu\xx=0$, $\cS^2\cX^\sigma\xx=0$.

\bigskip

In this approach we suppose that the ghost expansion term
$\cX^\mu\xx$  can be interpreted as a consequence of the connection extension $\Omega^{(\rho)}_{(\mu,\nu)}\xx$ term, and goes to zero when this one vanishes.
So we expand $\cX^\mu\xx$  in terms of the connection extension and its derivatives, with coefficients  higher order ghosts $\cC^{(\lambda_1,\cdots,
\lambda_{j})}\xx$ with $\Phi,\Pi$ charge equal to one, and we choose
\begin{eqnarray}
\cX^\sigma\xx&=&\sum_{j=1}^\infty\Biggl[\cC^{(\eta_1,\cdots,
\eta_{j})}\xx\Biggl(\cD_{\eta_1}\cdots
\cD_{\eta_{j-2}}\Omega_{(\eta_{j-1}\eta_j)}^\sigma\xx\Biggr)\Biggr]\nn\\
\lbl{cXexpansion}
\end{eqnarray}
We remark that the $\cC^{(\eta_1,\cdots,
\eta_{j})}\xx$ ghosts tensors order starts from two and goes to infinity.

At this stage it is evident that the Equation \rf{cXexpansion}
must be consistent with Equations \rf{SX}and \rf{somegacov}
which should provide the B.R.S. variations of
$\cC^{(\eta_1,\cdots,\eta_{j})}\xx$, which can be derived
picking out the terms $\Biggl(\cD_{\eta_1}\cdots
\cD_{\eta_{j-2}}\Omega_{(\eta_{j-1}\eta_j)}^\sigma\xx\Biggr)$,
 (in which is placed the free $\sigma$ index).

%Now, in the calculation of the B.R.S transformation of
%\rf{cXexpansion}, we have to keep our mind, if we want to
%extract these terms,   on the free index location.

This inspection is (almost) always successful, but encounter an
obstacle provided by the presence of the last term of the
$\Omega_{(\mu,\nu)}^\sigma\xx$ B.R.S transformations
($\cX^{(\lambda)}\xx\cR^{(\sigma)}_{([\lambda,\rho],\eta)}\xx$),in
Equation \rf{somegacov}, which exhibits, the free index placed
in the top of the curvature, and  {\it{not}} on the
$\cX^\lambda\xx$ term. So, taking into account Equation
\rf{cXexpansion}), the free index misplacement forbids the
complete extraction of these terms.

Indeed must be natural suppose that this extension process is not an easy trip, and the geometry 'a priori' puts barriers to this program.
 So, in order to bypass this obstacle, we have to fix  conditions, which would assure the success of the operation.

The most simple requirement is to fix (this is one of the key
points of the paper):
\begin{eqnarray}
&&\cX^{(\lambda)}\xx\cR^{(\mu)}_{([\lambda,\rho],\sigma)}\xx
%\equiv\sum_{j=1}^\infty\Biggl[\cC^{(\eta_1,\cdots,
%\eta_{j})}\xx\Biggl(\cD_{\eta_1},\cdots,
%\cD_{\eta_{j-2}}\Omega_{(\eta_{j-1}\eta_j)}^\lambda\xx\Biggr)\Biggr]\cR^{(\mu)}_{([\lambda,\rho],\sigma)}\xx
=
0
\lbl{Xconstraint}
\end{eqnarray}
\footnote{an alternative  second solution can be given choosing:
\begin{eqnarray}
\cX^{(\lambda)}\xx\cR^{(\mu)}_{([\lambda,\rho],\sigma)}\xx= \cX^{(\mu)}\xx\cG_{(\rho,\sigma)}\xx
\end{eqnarray}
which can be appealing in case of Einstein spaces }

\bigskip

 But, substituting Equation \rf{cXexpansion}  into Equation \rf{Xconstraint}we get:
\begin{eqnarray}
\sum_{j=1}^\infty\Biggl[\cC^{(\eta_1,\cdots,
\eta_{j})}\xx\Biggl(\cD_{\eta_1},\cdots,
\cD_{\eta_{j-2}}\Omega_{(\eta_{j-1}\eta_j)}^\lambda\xx\Biggr)\Biggr]\cR^{(\mu)}_{([\lambda,\rho],\sigma)}\xx
=
0
\lbl{Xconstraint1}
\end{eqnarray}

The constraint in Equation \rf{Xconstraint1} can be verified (
due to its Faddeev-Popov content) if we impose:
\begin{eqnarray}
&&\Biggl[\Biggl(\cD_{\eta_1}\cdots
\cD_{\eta_{j-2}}\Omega_{(\eta_{j-1}\eta_j)}^\lambda\xx\Biggr)\Biggr]\cR^{(\mu)}_{([\lambda,\rho],\sigma)}\xx=
\cM_{((\eta_1,\cdots,\eta_j),(\tau_1,\cdots,\tau_j);\rho,\lambda)}^{(\mu)}\xx\cC^{(\eta_1,\cdots,
\eta_{j})}\xx\nn\\
&&\cM_{((\eta_1,\cdots,\eta_j),(\tau_1,\cdots,\tau_j);\rho,\lambda)}^{(\mu)}\xx
=\cM_{(
(\tau_1,\cdots,\tau_j),(\eta_1,\cdots,\eta_j);\rho,\lambda)}^{(\mu)}\xx\nn\\
&&\forall j \geq 1\nn\\
\lbl{cXexpansion1}
\end{eqnarray}

We underline that Equation \rf{cXexpansion1}, if inverted, defines  the $\cC^{(\eta_1,\cdots,
\eta_{j})}\xx$ ghosts in term of the $\Omega_{(\rho,\sigma)}^\lambda\xx$ covariant derivatives. This fact has to be examined later  in all its details.

Before this we underline that,if the condition \rf{Xconstraint}
is fulfilled,  its B.R.S.
transformation too must be zero; that is:
\begin{eqnarray}
\cS\biggl(\cX^{(\lambda)}\xx\cR^{(\mu)}_{([\lambda,\rho],\sigma)}\xx\biggr)=0
\end{eqnarray}
so, taking into account Equation \rf{Xconstraint}, only the terms coming from the connection algebra extension survive, so we must require:
\begin{eqnarray}
\cX^{(\lambda)}\xx
\biggl(D_\lambda\Omega^{(\mu)}_{(\rho,\sigma)}\xx-\cD_\rho\Omega^{(\mu)}_{(\lambda,\sigma)}\xx\biggr)=0\nn\\
\lbl{extraterm}
\end{eqnarray}

A way out can be adopted  requiring:
\begin{eqnarray}
\Omega^{(\mu)}_{(\rho,\sigma)}\xx=\cD_\rho\cK^{(\mu)}_{(\sigma)}\xx
\lbl{Omegasolution}
\end{eqnarray}

so it is easy to verify,if we assume Equations \rf{Xconstraint} \rf{Omegasolution}:

\begin{eqnarray}
&&\cS\biggl(\cX^{(\lambda)}\xx\cR^{(\mu)}_{([\lambda,\rho],\sigma)}\xx\biggr)=\cX^{(\lambda)}\xx\biggl(\Biggl[D_\lambda,\cD_\rho\Biggr]\cK^{(\mu)}_{(\sigma)}\xx
\biggr)\nn\\
&&=\cX^{(\lambda)}\xx\biggl(\cR^{(\eta)}_{([\lambda,\rho],\sigma)}\xx\cK^{(\mu)}_{(\eta)}\xx
-\cR^{(\mu)}_{([\lambda,\rho],\eta)}\xx\cK^{(\eta)}_{(\sigma)}\xx\biggr)=0
\lbl{extraterm1}
\end{eqnarray}
(where we have neglect the terms coming from the non extended
diffeomorphism algebra, which due to Equation \rf{Xconstraint}
are zero)

Now from Equations \rf{somegacov} and \rf{Omegasolution} we can derive the $\cK^{(\mu)}_{(\sigma)}\xx$ B.R.S. variation:

\begin{eqnarray}
\cS\cK^{(\mu)}_{(\sigma)}\xx=\cC^{(\lambda)}\cD_{(\lambda)}\cK^{(\mu)}_{(\sigma)}\xx+
\cD_{(\sigma)}\cC^{(\lambda)}\cK^{(\mu)}_{(\lambda)}\xx-\cD_{(\lambda)}\cC^{(\mu)}
\cK^{(\lambda)}_{(\lambda)}\xx+\cD_{(\sigma)}\cX^{(\mu)}\xx-\cK_{(\sigma)}^{(\eta)}\xx\cK^{(\mu)}_{(\eta)}\xx
\lbl{sK}
\end{eqnarray}

So, defining
$\cM^{((\eta_1,\cdots,\eta_j),(\tau_1,\cdots,\tau_j);\rho,\lambda)}_{(\mu)}\xx$
such that:
\begin{eqnarray}
\cM_{((\eta_1,\cdots,\eta_j),(\tau_1,\cdots,\tau_j);\rho,\lambda)}^{(\mu)}\xx
\cM^{((\eta_1,\cdots,\eta_j),(\tau'_1,\cdots,\tau'_j);\rho',\lambda')}_{(\mu')}\xx=
\delta^{(\tau'_1,\cdots,\tau'_j)}_{(\tau_1,\cdots,\tau_j)}\delta^{\mu}_{\mu'}\delta^{\rho'}_{\rho}\delta^{\lambda'}_{
\lambda}
\end{eqnarray}
the Equation, after taking into account the Equation \rf{Omegasolution}, the  Equation\rf{cXexpansion1} can be inverted, and we obtain the
 expressions of the $\cC^{(\tau_1,\cdots,\tau_{j})}\xx$ in terms of the $\cK_\rho^\sigma\xx$ connection extension parameters and their covariant derivatives.
%which fix an infinite chain od covariant differential equations for $\cK_\rho^\sigma\xx$ in terms of the %ghosts $\cC^{(\tau_1,\cdots,\tau_{j})}\xx$ with coefficients the curvatures and %$\cM_{((\eta_1,\cdots,\eta_j),(\tau_1,\cdots,\tau_j);\rho,\lambda)}^{(\mu)}\xx$.
\begin{eqnarray}
\cC^{(\eta_1,\cdots,\eta_{j})}\xx=\cM^{((\eta_1,\cdots,\eta_j),(\tau_1,\cdots,\tau_j);\rho,\lambda)}_{(\mu)}\xx
\Biggl(\cD_{\tau_1},\cdots,
\cD_{\tau_{j-1}}\cK_{(\tau_j)}^\sigma\xx\Biggr)\cR^{(\mu)}_{([\sigma,\rho],\lambda)}\xx\nn\\
\end{eqnarray}
with this we have the parameters collapse, which, from an infinite multitude
 ($\cC^{(\eta_1,\cdots,\eta_{j})}\xx, j=1,\cdots,\infty$), are  reduced only to a couple ($\cC^{(\eta)}\xx,\cK^{(\rho)}\xx$).

\bigskip
So the breaking term $\cX^\sigma\xx$ is takes its final form:
\begin{eqnarray}
\cX^\sigma\xx&=&\sum_{j=1}^\infty\Biggl[
\cM^{((\eta_1,\cdots,\eta_j),(\tau_1,\cdots,\tau_j);\rho,\eta)}_{(\mu)}\xx
\Biggl(\cD_{\tau_1},\cdots,
\cD_{\tau_{j-1}}\cK_{(\tau_j)}^\lambda\xx\Biggr)\cR^{(\mu)}_{([\eta,\rho],\lambda)}\xx
\Biggl(\cD_{\eta_1},\cdots,
\cD_{\eta_{j-1}}\cK_{(\eta_j)}^\sigma\xx\Biggr)\Biggr]\nn\\
\lbl{cXexpansion2}
\end{eqnarray}

In the next Section we shall apply this extension to a new calculation of the Gravitational Anomalies\cite{Alvarez-Gaume:1983ig} .
We shall see that, within our treatment, the  $\cX^\sigma\xx$ term will play a secondary role, so we skip any further investigation on the $\cM^{((\eta_1,\cdots,\eta_j),(\tau_1,\cdots,\tau_j);\rho,\lambda)}_{(\mu)}\xx$ terms.

\section{A B.R.S. approach on the Characteristic Classes and Gravitational Anomalies  }

\lbl{anomalies}

We here rephrase  in this context, an approach of the present author\cite{Bandelloni:1988ws} to derive  invariants in an abstract way directly from the B.R.S. diffeomorphisms algebra.Then, a trick we shall see later, will permit to link them to the world of Field Theory.

We start from the study of the covariant ghost derivative \cite{Bandelloni:2012ze}:
\begin{eqnarray}
\cD_{(\mu)}\cC^{(\nu)}\xx\equiv \prt_{(\mu)}\cC^{(\nu)}\xx- \Gamma_{(\mu,\lambda)}^{(\nu)}\xx\cC^{(\lambda)}\xx
\end{eqnarray}

Its B.R.S. variation takes the form:

\begin{eqnarray}
\cS\cD_\nu\cC^\mu\xx =\biggl(\Lambda^\mu_\nu\xx+\cD_\nu\cC^\lambda\xx\cD_\lambda\cC^\nu\xx\biggr)
\lbl{sdc}
\end{eqnarray}
where:
\begin{eqnarray}
\Lambda_{(\mu)}^{(\nu)}\xx=\frac{1}{2}\cC^{(\rho)}\xx\cC^{(\sigma)}\xx
\cR^{(\nu)}_{([\rho,\sigma],\mu)}\xx+\cD_\mu\cX^\mu+\cC^\lambda\xx\Omega_{(\lambda,\mu)}^\nu\xx
\end{eqnarray}
Its variation  is astonishing simple :
\begin{eqnarray}
\cS \Lambda^\mu_\nu\xx=\cD_\nu\cC^\lambda\xx\Lambda_\lambda^\mu\xx-\Lambda_\nu^\lambda\xx\cD_\lambda\cC^\mu\xx \equiv{\Biggl[\cD\cC\xx,\Lambda\xx\Biggr]}^\mu_\nu
\lbl{slambda}
\end{eqnarray}

The fundamental result we get from Equation \rf{slambda} is:
\begin{eqnarray}
\cS Tr{\Biggl[\Lambda^n\xx\Biggr]}=0
\lbl{invariance0}
\end{eqnarray}
for all $n \geq 1${\footnote{For sake of brevity we introduce the notations:
\begin{eqnarray}
{\biggl({\cD\cC\xx}^m\biggr)}_\nu^\mu\xx\equiv{\biggl(\cD_\nu\cC^{\rho_1}\xx\cD_{\rho_1}\cC^{\rho_2}\xx\cdots
\cD_{\rho_m}\cC^\mu\xx\biggr)}
\end{eqnarray}

\begin{eqnarray}
{\biggl({\Lambda\xx}^n\biggr)}_\nu^\mu\xx\equiv{\biggl(\Lambda_\nu^{\sigma_1}\xx\Lambda_{\sigma_1}
^{\rho_2}\xx\cdots
\Lambda_{\sigma_n}^\mu\xx\biggr)}
\end{eqnarray}}}
So we get an infinity of  local invariant under  B.R.S.diffeomorphism.
It is obvious to realize that this property is verified even if the infinitesimal transformations are extended or not.
In the case of usual diffeomorphisms we obtain:
\begin{eqnarray}
\zer{\Lambda}_{(\mu)}^{(\nu)}\xx=\frac{1}{2}\cC^{(\rho)}\xx\cC^{(\sigma)}\xx
\cR^{(\nu)}_{([\rho,\sigma],\mu)}\xx
\end{eqnarray}

and the invariants provide a new B.R.S. way of thinking of the Chern characteristic  classes.
The present strategy allows to go beyond the usual linear transformations and to explore what is next to the habitual infinitesimal border.

In our framework, the most compelling question is that these invariants are cohomology elements or not.

We shall analyze this problem, constructing the solutions for each $n$, as coboundary polynomials,with specific and particular properties.

%\section{ $Tr\biggl[\Lambda^{n)} \xx\biggr]$ B.R.S decompositions, Anomalies construction and %management}

First of all we state here the main  result (whose detailed
calculations are reported in the Appendix \rf{appendix}) we
shall use here:
\begin{Results}
The term $Tr\Biggl[\Lambda^{(n)}\xx\Biggr]$ is a total coboundary and take the general expression:

\begin{eqnarray}
Tr\Biggl[\Lambda^{(n)}\xx\Biggr]
&=&\cS Tr \widehat{\Sigma}^{(2n-1)}\xx\nn\\
\widehat{\Sigma}^{(2n-1)}\xx&\equiv&
\sum_{r=0}^{(n-1)}  Tr \Biggl[\cD\cC\xx\Biggl\{\Biggl(-{\biggl(\cS\cD\cC\xx\biggr)}^{(n-1-r)}+\Lambda^{(n-1-r)}\xx\biggr)
\Psi^{(r)}\xx\Biggr)\nn\\
&+&{\frac{1}{(2r+1)}\biggl({\biggl(\cD\cC\xx\cD\cC\xx\biggr)}^{(r)}\biggl(\cS\cD\cC\xx \biggr)}^{(n-r-1)}
\biggr)\Biggr\}\Biggr]\nn\\
%&\equiv&\
\lbl{result}
\end{eqnarray}
where:
\begin{eqnarray}
\Psi^{(0)}\xx&=& 1\nn\\
\Psi^{(1)}\xx&=& (\cD\cC\xx\cD\cC\xx)\nn\\
\Psi^{(2)}\xx&=& \biggl(\cD\cC\xx\cS(\cD\cC\xx\cD\cC\xx)+{(\cD\cC\xx\cD\cC\xx)}^2\biggr)\nn\\
&=&\cdots\nn\\
\Psi^{(r)}\xx&=&\cD\cC\xx\cS\Psi^{(r-1)}\xx+{(\cD\cC\xx\cD\cC\xx)}^{(r)}\nn\\
r&=&1,\cdots n,\qquad\cS\Psi^{(0)}\xx=0\nn\\
&&\Phi\Pi\Biggl[\Psi^{(r)}\xx\Biggr]=2r\nn\\
\end{eqnarray}

\end{Results}

The calculations which lead to the previous conclusions are
reported in the Appendix \rf{appendix}.

With these results we shall be able to construct the gravitational anomalies in all even dimensions.

we give some result for the lowest value:
\begin{eqnarray}
Tr{\Biggl[\Lambda^2\xx\Biggr]} &=&\cS Tr\Biggl(\cD\cC\xx\Lambda\xx+\frac{1}{3}\cD\cC\xx\cD\cC\xx\cD\cC\xx\Biggr)\nn\\
\nn\\
Tr{\Biggl[\Lambda^3\xx\Biggr]}&=&\cS Tr\Biggl(\cD\cC\xx\Lambda^2\xx+\frac{1}{2}\cD\cC\xx\cD\cC\xx\cD\cC\xx\Lambda\xx
-\frac{1}{10}\cD\cC\xx\cD\cC\xx\cD\cC\xx\cD\cC\xx\cD\cC\xx\Biggr)\nn\\
Tr{\Biggl[\Lambda^4\xx\Biggr]} &=&\cS Tr \Biggl(\cD\cC\xx\Lambda^3\xx+ +\frac{1}{5}\cD\cC\xx\Lambda\xx\cD\cC\xx\cD\cC\xx\Lambda\xx+ \frac{2}{5}\cD\cC\xx\cD\cC\xx\cD\cC\xx\Lambda^2\xx\nn\\&+&\frac{1}{5}\cD\cC\xx\cD\cC\xx\cD\cC\xx\cD\cC\xx\cD\cC\xx\Lambda\xx
-\frac{1}{35}\cD\cC\xx\cD\cC\xx\cD\cC\xx\cD\cC\xx\cD\cC\xx\cD\cC\xx\cD\cC\xx\Biggr)\nn\\
\nn\\
\end{eqnarray}

The previous results are general and hold for all $\cX^\mu\xx$ and $\Omega_{(\mu,\nu)}^\sigma\xx$ which obey the previous B.R.S transformations, that is for whatever consistent extension.

We have seen in  past\cite{Bandelloni:1988ws}, that with algebraic techniques, writing the common derivative operator in the Fock space\cite{Dixon:1979bs}, in term of B.R.S ghosts and operator:,
\begin{eqnarray}
\prt_{\mu}=\Biggl\{\delta_{B.R.S.},\frac{\prt}{\prt \cC^{(\mu)}\xx}\Biggr\}\nn\\
\lbl{prt}
\end{eqnarray}
(where $\delta_{B.R.S.}$ is the B.R.S. functional operator) {\it{it is possible to relate local B.R.S. totally invariant polynomials, to invariant objects modulo total derivatives\cite{Bandelloni:2012ze}, and to derive all the  elements of the so called "descent Equations"\cite{Stora:1983ct}.}}
We remark that the highest underived $\cC^\mu\xx$ ghost content of the previous object is the local parallel term of the invariant $Tr\biggl[{\biggl(\cC^\mu\xx\cC^\nu\xx \cR^{\sigma}_{([\mu,\nu],\rho)}\xx\biggr)}^n\biggr]$, so
 our trick allows to extend the idea of the so-called "Russian formula" idea in presence of extensions too, and to write in a general way Topological Actions and Anomalies of the extended symmetries in fixed dimensions. So, if $\Delta^{q,\natural}_{j}\xx$ are the total local invariants with Faddeev- Popov ($\Phi,\Pi$) charge equal to $q$ and form degree equal to $j$, then the invariants (modulo total derivatives) with $\Phi,\Pi$ charge $p$ and degree n takes the general expression\cite{Bandelloni:1988ws}:
 \begin{eqnarray}
 \Delta^{p}_n \xx= \Delta^{p,\natural}_n \xx +\sum_{j=1}^n (-1)^j \frac{1}{j!} dx^1\wedge\cdots\wedge dx^j \frac{\prt^j\Delta^{p+j,\natural}_{d-j}\xx}{\prt \cC^{\mu_1}\xx\cdots\prt\cC^{\mu_j}\xx} +{\bf d}\widehat{\Delta}^{p}_{n-1} \xx+\cS\widehat{\Delta}_n^{p-1}\xx
\end{eqnarray}

So the more direct application of this  approach, is the direct calculations of Topological Actions and Anomalies of models whose local symmetries are the reparametrizations:

So, if we calculate  the Lagrangian anomalies\cite{Alvarez-Gaume:1983ig}, from the previous coboundary terms,
 we derive, besides the usual term, a further one whose $\Phi,\Pi$ charge is carried by the $\Omega_{(\mu,\nu)}^\sigma\xx$ ghost term, and we have to remark that
 {\it{,in any dimension, the anomaly is independent from the $\cX^\mu\xx$ invariant
 contributions, since in Eq \rf{cXexpansion} expansion no first order underived $\cC^\mu\xx$
 ghost is present}}. This fact justifies the discussion cut we have performed at the end of the Section \rf{algebraextension},
 and will have  relevance in the following.

For example in the two dimensional case, the anomaly takes the form:
\begin{eqnarray}
\Delta^\natural_{[\mu,\nu]} \xx dx^\mu\wedge dx^\nu &=&
\frac{1}{2} dx^\mu\wedge dx^\nu \frac{\prt}{\prt \cC^\mu\xx}\frac{\prt}{\prt \cC^\nu\xx}\Biggl(Tr\Biggl(\cD\cC\xx(\frac{1}{2}\cC\xx\cC\xx\cR\xx+\cC\xx\cD\cK\xx+\cD\cX\xx)\nn\\
&+&\frac{1}{3}\cD\cC\xx\cD\cC\xx\cD\cC\xx\Biggr)\Biggr)\nn\\
\end{eqnarray}

and we derive:

\begin{eqnarray}
\Delta_{[\mu,\nu]} \xx dx^\mu\wedge dx^\nu  = \epsilon^{\mu,\nu}\xx\Biggl(-\cD_\rho \cC^\sigma \xx+\cK^\sigma_\rho\xx
\Biggr)\cR^{\rho}_{([\mu,\nu],\sigma)}\xx d x^2\nn\\
\lbl{2dimanomalycov}
\end{eqnarray}
where we have adopted the full covariant way adopted in the Reference \cite{Bardeen:1984pm}.

So if we embed this anomaly within a B.R.S. Lagrangian framework with  an improved quantum Topological Action  $\Gamma$ , the Quantum Action Principle\cite{Lam:1973qa} fix this anomaly as:
\begin{eqnarray}
\delta_{B.R.S} \Gamma =\int \Delta_{[\mu,\nu]} \xx dx^\mu\wedge dx^\nu
\end{eqnarray}

so, introducing the Ward operators:
\begin{eqnarray}
\cW_\mu\xx\equiv \biggl\{\frac{\prt}{\prt\cC^\mu\xx},\delta_{B.R.S}\biggr\};\qquad\cW_\mu^\sigma\xx\equiv \biggl\{\frac{\prt}{\prt\cK_\sigma^\mu\xx},\delta_{B.R.S}\biggr\}\nn\\
\end{eqnarray}
the previous calculations give:
\begin{eqnarray}
\cW_\mu\xx\Gamma=\cD_\sigma\cW_\mu^\sigma\xx\Gamma
\end{eqnarray}
at the anomaly level.

So it is possible to define an improved local Ward operator ${\widetilde{\cW}}_\mu\xx\equiv\Biggl(\cW_\mu\xx-\cD_\sigma\cW_\mu^\sigma\xx\Biggr)$  such that it is possible to have
\begin{eqnarray}
{\widetilde{\cW}}_\mu\xx\Gamma=0
\lbl{compensation}
\end{eqnarray}
for which a symmetry is locally restored (at the anomaly quantum level).

The same conclusion is achieved going to four dimensions, since in this case we get:
\begin{eqnarray}
&&\Delta^\natural_{[\mu,\nu, \rho,\sigma]} \xx dx^\mu\wedge dx^\nu \wedge dx^\rho\wedge dx^\sigma\nn\\
&&=
\frac{1}{4!} dx^\mu\wedge dx^\nu\wedge dx^\rho\wedge dx^\sigma \frac{\prt}{\prt \cC^\mu\xx}\frac{\prt}{\prt \cC^\nu\xx}\frac{\prt}{\prt \cC^\rho\xx}\frac{\prt}{\prt \cC^\sigma\xx}
\Biggl(Tr\Biggl(\cD\cC\xx(\frac{1}{2}\cC\xx\cC\xx\cR\xx+\cC\xx\Omega\xx+\cD\cX\xx)\nn\\
&& (\frac{1}{2}\cC\xx\cC\xx\cR\xx+\cC\xx\Omega\xx+\cD\cX\xx)
+\frac{1}{2}\cD\cC\xx\cD\cC\xx\cD\cC\xx(\frac{1}{2}\cC\xx\cC\xx\cR\xx+\cC\xx\Omega\xx+\cD\cX\xx)\nn\\&&
-\frac{1}{10}\cD\cC\xx\cD\cC\xx\cD\cC\xx\cD\cC\xx\cD\cC\xx\Biggr)\Biggr)
\end{eqnarray}
and, after performing all the calculations,the anomaly can be rewritten, as well known, in a full covariant way with the aid of total derivatives and coboundary terms,as :
\begin{eqnarray}
&&\Delta^\natural_{[\mu,\nu, \rho,\sigma]} \xx dx^\mu\wedge dx^\nu \wedge dx^\rho\wedge dx^\sigma
=
\epsilon^{\mu,\nu,\rho,\sigma}\Biggl(-\cD_\lambda\cC^\eta\xx+\cK^\eta_\lambda\xx\Biggr)
\cR^\tau_{([\mu,\nu],\eta)}\xx\cR^\lambda_{([\rho,\sigma],\tau])}\xx
d^4 x\nn\\
\lbl{4dimanomalycov}
\end{eqnarray}

and the same result as before  in Equation \rf{compensation} is achieved.

We can generalize this result within our approach, since the solution of our problem is revealed   noting that

 \begin{eqnarray}
 \cS\Biggl(\cD_\nu\cC^\mu\xx-\cK^\mu_\nu\xx\Biggr)=
 \Biggl(\cD_\nu\cC^\lambda\xx-\cK^\lambda_\nu\xx\Biggr)\Biggl(\cD_\lambda\cC^\mu\xx-\cK^\mu_\lambda\xx\Biggr)+\zer{\Lambda}_\nu^\mu\xx
 \lbl{conds}
 \end{eqnarray}

 The previous equation \rf{conds} is the direct evidence of our
 results.
This rewrites Equation \rf{sdc} putting all the extensions
inside the new ghost term $\cK^\mu_\nu$,
  %which shows a parallel running together with the covariant derivatives $\cD_\nu\cC^\mu\xx$,
  leading to a compensation mechanism  which produces the invariants of the non-extended algebra.
  %But in this construction the ghosts structure has been deformed.

This is the reason why in the two and four dimensional anomalies in Equations \rf{2dimanomalycov}  and \rf{4dimanomalycov} we find the well-known covariant gravitational anomalies  \cite{Bardeen:1984pm} \cite{Ader:1990ex} with the replacement $\cD_\nu\cC^\mu\xx\lra\cD_\nu\cC^\mu\xx-\cK^\mu_\nu\xx$.

With this result, the temptation to adopt a cancellation
procedure, is immediate.

As a final investigation, one may ask what is the   Classical  Topological Action $\Gamma^{Classical}$ invariant under the full extended symmetry;
so at the Classical limit we have to verify
\begin{eqnarray}
&&\cW_\mu\xx\Gamma^{Classical}=0\nn\\
&&\cW_\mu^\sigma\xx\Gamma^{Classical}=0\nn\\
\lbl{Classical}
\end{eqnarray}

 This means that the Action is a topological function of the curvature, invariant under usual reparametrizations, and, separately, their extensions in the $\cK^{(\mu)}_{(\nu)}\xx$ ghosts.

Now, it is easy to derive from Equations \rf{sR} , \rf{Omegasolution}  and \rf{extraterm1} that:
\begin{eqnarray}
&&\cW^\tau_\eta\yy \cR^\mu_{([\lambda,\rho],\nu)}\xx=\Biggl(\cR^\tau_{([\lambda,\rho],\nu)}\xx\delta^\mu_\eta-\cR^\mu_{([\lambda,\rho],\eta)}\xx\delta^\tau_\nu\Biggr)\delta(x-y)\nn\\
\end{eqnarray}

which fix the Action as a function of   $ \cR^\eta_{([\mu,\nu],\eta)}\xx $; such as, for example, in $2n$ dimensions the Lagrangian takes
the form: $\cL^{Classical}\xx= \epsilon^{(\mu_1,\mu_2,\cdots\mu_{2n-1},\mu_{2n})} \cR^{\eta_1}_{([\mu_1,\mu_2],\eta_1)}\xx\cdots\cR^{\eta_n}_{([\mu_{2n_1},\mu_{2n}],\eta_n)}\xx$.

\section{Conclusions}
\lbl{conclusions}
 The results we have presented here are very general.Our
investigation takes origin only from the curiosity to see
beyond a smooth reparametrization, and we make only use of
algebraic consistency and stability (B.R.S. nilpotency).The
message we derive is that sometimes, just round the corner, we
can find the solutions of our troubles.Condition
\rf{Xconstraint} is fundamental for this symmetry, not only it
shows the right lane which the expansion $\cX^\mu\xx$ term
(from which the anomaly in independent) has to go through, but
also reduces the ghosts from infinity to two.

 After this shrinking the algebra produces a new symmetry, embodied in condition \rf{conds}, which reconduces the problem to a quiet normality.

 Obviously the solution might not to be unique: this gives new forces to the study of the non linear extensions of algebras.

\appendix
\section{Appendix}
\lbl{appendix}

 In this Appendix we report the calculations which give the
result in
 \rf{result}.

 Essentially, this comes from the iterations of the formula:

\begin{eqnarray}
&&\Lambda^{(r+1)}\xx=\cS\biggl(\Lambda^{(r)}\xx\cD\cC\xx\biggr)
-\cD\cC\xx\Lambda^{(r)}\xx\cD\cC\xx\nn\\
&&r\geq 0\nn\\
\end{eqnarray}
where we have defined $\Lambda^{(0)}\xx=1$.

The first iteration, gives:
\begin{eqnarray}
&&Tr\Biggl[\Lambda^{(n)}\xx\Biggr]=\cS Tr\Biggl[\biggl(\Lambda^{(n-1)}\xx\cD\cC\xx\biggr)\Biggr]- Tr\Biggl[\cD\cC\xx\Lambda^{(n-1)}\xx\cD\cC\xx\Biggr]\nn\\
\end{eqnarray}

The process  goes on as:
\begin{eqnarray}
&=&\cS Tr\Biggl[\biggl(\Lambda^{(n-1)}\xx\cD\cC\xx\biggr)\Biggr]+ Tr\Biggl[\cD\cC\xx\cD\cC\xx\Lambda^{(n-1)}\xx\Biggr]\nn\\
&=&\cS Tr\Biggl[\biggl(\Lambda^{(n-1)}\xx\cD\cC\xx\biggr)\Biggr]+ Tr\Biggl[\cD\cC\xx\cD\cC\xx\biggl(\cS\biggl(\Lambda^{(n-2)}\xx\cD\cC\xx\biggr)
-\cD\cC\xx\Lambda^{(n-2)}\xx\cD\cC\xx\biggr)\Biggr]\nn\\
&=&\cS Tr\Biggl[\biggl(\Lambda^{(n-1)}\xx\cD\cC\xx\biggr)+\biggl((\cD\cC\xx\cD\cC\xx)\Lambda^{(n-2)}\xx\cD\cC\xx\biggr)\Biggr]
\nn\\&-&Tr\Biggl[\cS(\cD\cC\xx\cD\cC\xx)\Lambda^{(n-2)}\xx\cD\cC\xx\Biggr]
-Tr \Biggl[(\cD\cC\xx\cD\cC\xx)\cD\xx\Lambda^{(n-2)}\xx\cD\cC\xx\Biggr]\nn\\
%\end{eqnarray}
%\begin{eqnarray}
&=&\cS Tr\Biggl[\biggl(\Lambda^{(n-1)}\xx\cD\cC\xx\biggr)
+
\biggl((\cD\cC\xx\cD\cC\xx)\Lambda^{(n-2)}\xx\cD\cC\xx\biggr)\Biggr]
\nn\\
&+&Tr\Biggl[\biggl(\cD\cC\xx\cS(\cD\cC\xx\cD\cC\xx)+{(\cD\cC\xx\cD\cC\xx)}^2\biggr)\Lambda^{(n-2)}\xx\Biggr]
\nn\\
&=&\cS Tr\Biggl[\biggl(\Lambda^{(n-1)}\xx\cD\cC\xx\biggr)
+
\biggl((\cD\cC\xx\cD\cC\xx)\Lambda^{(n-2)}\xx\cD\cC\xx\biggr)\Biggr]
\nn\\
&+&Tr\Biggl[\biggl(\cD\cC\xx\cS(\cD\cC\xx\cD\cC\xx)+{(\cD\cC\xx\cD\cC\xx)}^2\biggr)
\Biggl(\cS\biggl(\Lambda^{(n-3)}\xx\cD\cC\xx\biggr)
-\cD\cC\xx\Lambda^{(n-3)}\xx\cD\cC\xx
\Biggr)
\Biggr]\nn\\
&=&\cdots\nn\\
\end{eqnarray}
%\begin{eqnarray}

so we argue a first general result:
\begin{eqnarray}
Tr\Biggl[\Lambda^{(n)}\xx\Biggr]&=& Tr\cS\Biggl[\sum_{r=0}^{n-1} \Psi^{(r)}\xx \Lambda^{(n-1-r)}\xx\cD\cC\xx\Biggr] +Tr \Biggl[\Psi^{(n)}\xx \Biggr]
\end{eqnarray}

where:
\begin{eqnarray}
\Psi^{(0)}\xx&=& 1\nn\\
\Psi^{(1)}\xx&=& (\cD\cC\xx\cD\cC\xx)\nn\\
\Psi^{(2)}\xx&=& \biggl(\cD\cC\xx\cS(\cD\cC\xx\cD\cC\xx)+{(\cD\cC\xx\cD\cC\xx)}^2\biggr)\nn\\
&=&\cdots\nn\\
\Psi^{(r)}\xx&=&\cD\cC\xx\cS\Psi^{(r-1)}\xx+{(\cD\cC\xx\cD\cC\xx)}^{(r)}\nn\\
r&=&1,\cdots n\nn\\
\end{eqnarray}

At this stage we are left here to show that  the $Tr \Biggl[\Psi^{(n)}\xx \Biggr]$ is a coboundary, and a new iteration  task must be accomplished.
\begin{eqnarray}
Tr \Biggl[\Psi^{(n)}\xx \Biggr]&=&Tr\Biggl[\cD\cC\xx\cS\Psi^{(n-1)}\xx+{(\cD\cC\xx\cD\cC\xx)}^{(n)}\Biggr]\nn\\
&=&Tr\Biggl[\cD\cC\xx\cS\Psi^{(n-1)}\xx\Biggr]\nn\\
&=&-\cS Tr\Biggl[\cD\cC\xx\Psi^{(n-1)}\xx\Biggr]+\Biggl[\biggl(\cS\cD\cC\xx\biggr)\Psi^{(n-1)}\xx\Biggr]\nn\\
&=& -\cS Tr\Biggl[\cD\cC\xx\Psi^{(n-1)}\xx\Biggr]+Tr\Biggl[\biggl(\cS\cD\cC\xx\biggr)
\biggl(\cD\cC\xx\cS\Psi^{(n-2)}\xx+{(\cD\cC\xx\cD\cC\xx)}^{(n-1)}\Biggr]\nn\\
&=&\cS Tr
\Biggl[-\cD\cC\xx\biggl(\Psi^{(n-1)}\xx+\biggl(\cS\cD\cC\xx\biggr)\Psi^{(n-2)}\xx\biggr)
+\frac{1}{(2n-1)}\biggl(\cD\cC\xx{(\cD\cC\xx\cD\cC\xx)}^{(n-1)}\biggr)\Biggr]\nn\\
&+&
Tr \Biggl[{\biggl(\cS\cD\cC\xx\biggr)}^2\Psi^{(n-2)}\xx\Biggr]\nn\\
&=&\cS Tr
\Biggl[\cD\cC\xx\biggl(\Psi^{(n-1)}\xx+\biggl(\cS\cD\cC\xx\biggr)\Psi^{(n-2)}\xx\biggr)
+\frac{1}{(2n-1)}\biggl(\cD\cC\xx{\biggl(\cD\cC\xx\cD\cC\xx\biggr)}^{(n-1)}\biggr)\Biggr]\nn\\
&+&
Tr \Biggl[{\biggl(\cS\cD\cC\xx\biggr)}^2\Biggl(\cD\cC\xx\cS\Psi^{(n-3)}\xx+{(\cD\cC\xx\cD\cC\xx)}^{(n-2)}\Biggr)\Biggr]\nn\\
&=&\cS Tr
\Biggl[-\cD\cC\xx\biggl(\Psi^{(n-1)}\xx+\biggl(\cS\cD\cC\xx\biggr)\Psi^{(n-2)}\xx
+{\biggl(\cS\cD\cC\xx\biggr)}^{2}\Psi^{(n-3)}\xx\biggr)
\nn\\
&+&\frac{1}{(2n-1)}\biggl(\cD\cC\xx{\biggl(\cD\cC\xx\cD\cC\xx\biggr)}^{(n-1)}\biggr)
+{(\cS\cD\cC\xx)}\frac{1}{(2n-3)}\biggl(\cD\cC\xx{\biggl(\cD\cC\xx\cD\cC\xx\biggr)}^{(n-2)}\biggr)\Biggr]\nn\\
&+&
Tr \Biggl[{\biggl(\cS\cD\cC\xx\biggr)}^2\Biggl(\cD\cC\xx\Psi^{(n-3)}\xx
\Biggr)\Biggr]\nn\\
&=& \cdots\nn\\
&=&\sum_{r=0}^{(n-1)} \cS Tr \Biggl[\cD\cC\xx\Biggl(-{\biggl(\cS\cD\cC\xx\biggr)}^{(n-1-r)}\Psi^{(r)}\xx\biggr)\Biggr)\nn\\
&+&{\biggl(\cS\cD\cC\xx \biggr)}^{(n-1-r)}\frac{1}{(2r+1)}\biggl({\biggl(\cD\cC\xx\cD\cC\xx\biggr)}^{(r)}
\biggr)\Biggr]\nn\\
\end{eqnarray}
where we have used:
\begin{eqnarray}
%&&Tr\Biggl[{(\cD\cC\xx\cD\cC\xx)}^{(n)}\Biggr]=0\nn\\
&&Tr\cS\Biggl[\biggl({(\cD\cC\xx\cD\cC\xx)}^{(n-r)} \biggr)\cD\cC\xx\Biggr]
=(2(n-r)+1)Tr\Biggl[\biggl(\cS\cD\cC\xx\biggr){\biggl(\cD\cC\xx\cD\cC\xx\biggr)}^{(n-r)}\Biggr]\nn\\
\end{eqnarray}
 and the iterations ends with:
 \begin{eqnarray}
&&Tr\Biggl[{(\cD\cC\xx\cD\cC\xx)}^{(n)}\Biggr]=0\nn\\
%&&Tr\cS\Biggl[\biggl({(\cD\cC\xx\cD\cC\xx)}^{(n-r)} \biggr)\cD\cC\xx\Biggr]
%=(2(n-r)+1)Tr\Biggl[\biggl(\cS\cD\cC\xx\biggr){\biggl(\cD\cC\xx\cD\cC\xx\biggr)}^{(n-r)}\Biggr]\nn\\
\end{eqnarray}
so our final result in Equation \rf{result} is obtained.
\subsection{Acknowledgement}
I am indebted  to Nicola Maggiore for discussions and  enlightening suggestions.	

%\bibliographystyle{unsrt}
%\bibliography{bibfiles}

\begin{thebibliography}{10}

\bibitem{DeWitt:1965jb} Bryce~S. DeWitt.
\newblock {Dynamical theory of groups and fields}.
\newblock {\em Conf.Proc.}, C630701:585--820, 1964.

\bibitem{Regge:1961px} T.~Regge.
\newblock {GENERAL RELATIVITY WITHOUT COORDINATES}.
\newblock {\em Nuovo Cim.}, 19:558--571, 1961.

\bibitem{Loll:1998aj} Renate Loll.
\newblock {Discrete approaches to quantum gravity in four-dimensions}.
\newblock {\em Living Rev.Rel.}, 1:13, 1998.

\bibitem{Zamolodchikov:1985wn} A.~B. Zamolodchikov.
\newblock Infinite additional symmetries in two-dimensional conformal quantum
  field theory.
\newblock {\em Theor. Math. Phys.}, 65:1205--1213, 1985.

\bibitem{DiFrancesco:1990qr} P.~Di~Francesco, C.~Itzykson, and
    J.B. Zuber.
\newblock {Classical W algebras}.
\newblock {\em Commun.Math.Phys.}, 140:543--568, 1991.

\bibitem{Becchi:1975nq} C.~Becchi, A.~Rouet, and R.~Stora.
\newblock Renormalization of gauge theories.
\newblock {\em Annals Phys.}, 98:287--321, 1976.

\bibitem{Alvarez-Gaume:1983ig} Luis Alvarez-Gaume and Edward
    Witten.
\newblock Gravitational anomalies.
\newblock {\em Nucl. Phys.}, B234:269, 1984.

\bibitem{Bandelloni:1988ws} G.~Bandelloni.
\newblock Diffeomorphism cohomology in quantum field theory models.
\newblock {\em Phys. Rev.}, D38:1156--1168, 1988.

\bibitem{Bandelloni:2012ze} Giuseppe Bandelloni.
\newblock {Nonlinear extensions of the reparametrization algebra: Algebraic
  construction, invariants and anomalies}.
\newblock {\em Int.J.Geom.Meth.Mod.Phys.}, 09:1250060, 2012.

\bibitem{Dixon:1979bs} J.~A. Dixon.
\newblock Cohomology and renormalization of gauge theories. 2.
\newblock HUTMP 78/B64.

\bibitem{Stora:1983ct} Raymond Stora.
\newblock Algebraic structure and topological origin of anomalies.
\newblock Seminar given at Cargese Summer Inst.: Progress in Gauge Field
  Theory, Cargese, France, Sep 1-15, 1983.

\bibitem{Bardeen:1984pm} William~A. Bardeen and Bruno Zumino.
\newblock {Consistent and Covariant Anomalies in Gauge and Gravitational
  Theories}.
\newblock {\em Nucl.Phys.}, B244:421, 1984.

\bibitem{Lam:1973qa} Yuk-Ming~P. Lam.
\newblock Equivalence theorem on bogolyubov-parasiuk-hepp-zimmermann
  renormalized lagrangian field theories.
\newblock {\em Phys. Rev.}, D7:2943--2949, 1973.

\bibitem{Ader:1990ex} Jean~Pierre Ader, Francois Gieres, and
    Yves Noirot.
\newblock {Gauged BRST symmetry and covariant gravitational anomalies}.
\newblock {\em Phys.Lett.}, B256:401--406, 1991.

\end{thebibliography}

\end{document}